\documentclass[conference]{IEEEtran}
\usepackage[utf8]{inputenc}
\usepackage[T1]{fontenc}
\usepackage{times}
\usepackage{graphicx}
\usepackage{booktabs}
\usepackage{multirow}
\usepackage{makecell}
\usepackage{xcolor}
\usepackage{hyperref}
\usepackage{amsmath}
\usepackage{amssymb}
\usepackage{caption,subcaption}
\usepackage{enumitem}
\usepackage{balance}
\usepackage{xspace}
\usepackage{listings}
\usepackage{colortbl}
\usepackage{pifont}
\usepackage{adjustbox}
\usepackage{diagbox}

\newcommand{\cmark}{\ding{51}}
\newcommand{\xmark}{\ding{55}}

\IEEEoverridecommandlockouts

\begin{document}

\title{Trust Me, Import This: Dependency Steering Attacks via Malicious Agent Skills}

\author{
    Yiyong Liu\textsuperscript{1\textcolor{blue!60!green}{$\ast$}}\thanks{\textcolor{blue!60!green}{$^\ast$}The first two authors made equal contributions.} \qquad
    Chia-Yi Hsu\textsuperscript{2\textcolor{blue!60!green}{$\ast$}} \qquad
    Chun-Ying Huang\textsuperscript{2} \\[0.15cm]
    Michael Backes\textsuperscript{1} \qquad
    Rui Wen\textsuperscript{3\textcolor{blue!60!green}{$\dag$}}\thanks{\textcolor{blue!60!green}{$^\dag$}Corresponding authors.} \qquad
    Chia-Mu Yu\textsuperscript{2\textcolor{blue!60!green}{$\dag$}} \\[0.3cm]
    \textit{\textsuperscript{1}CISPA Helmholtz Center for Information Security} \\
    \textit{\textsuperscript{2}National Yang Ming Chiao Tung University} \\
    \textit{\textsuperscript{3}Institute of Science Tokyo}
}

\maketitle

\begin{abstract}
LLM-powered coding agents increasingly make software supply chain decisions. They generate imports, recommend packages, and write installation commands. Prior work showed that these systems can hallucinate non-existent package names, which attackers may register as malicious packages. In this paper, we show that this risk is not only a passive model failure. It can be actively induced through the persistent Skill artifact.
We introduce Dependency Steering, an attack paradigm in which a malicious Skill biases a coding agent toward an attacker-controlled package during benign coding tasks. The attack does not require modifying model weights, training data, or user prompts.
To construct realistic attacks, we design a Skill-level optimization method that searches for localized semantic edits that preserve the apparent purpose of the original Skill while increasing targeted package generation. Across multiple coding-oriented LLMs and programming benchmarks, Dependency Steering achieves high targeted hallucination rates, transfers across models and task domains, and remains difficult for evaluated Skill scanners and LLM-based auditors to detect. Our results show that persistent agent instructions form an underexplored software supply chain attack surface.
\end{abstract}

\section{Introduction}
Modern coding agents increasingly make software supply chain decisions.
They no longer only complete code snippets, but also choose libraries, generate imports, modify dependency files, suggest installation commands, and interact with package ecosystems~\cite{repairagent,bouzenia2025,sweagent}.
Systems such as GitHub Copilot~\cite{github_copilot}, Cursor~\cite{cursor}, Claude Code~\cite{claude_code}, OpenAI Codex~\cite{openai_codex}, and Gemini Code Assist~\cite{gemini_code_assist} are therefore becoming part of the dependency selection process in everyday software development.
As developers rely on these agents for routine programming tasks, an incorrect or manipulated dependency recommendation can directly affect the software supply chain.

A concrete example of this risk is \emph{package hallucination}.
Prior work has shown that LLMs may recommend or import non-existent packages during code generation~\cite{pkghallu}.
Attackers can exploit this behavior by registering hallucinated package names in public registries and waiting for future users to install them, thereby turning model errors into supply chain compromises~\cite{neupane2023,zimmermann}.
Existing studies largely treat package hallucination as a passive model failure caused by imperfect memorization, probabilistic generation, or distributional uncertainty~\cite{pkghallu,surveyhallu_huang,surveyhallu_ji,krishna2025importingphantomsmeasuringllm,praticalhallu}.
This leaves an important question unanswered: can an attacker actively influence which dependencies a coding agent recommends?

In this paper, we show that dependency generation can be actively steered through persistent instruction artifacts.
We introduce \emph{Dependency Steering}, an attack paradigm in which a malicious Skill biases an LLM coding agent toward attacker-chosen packages during otherwise benign programming tasks.
Unlike passive package hallucination attacks, where the attacker waits for a model to naturally generate a vulnerable package name, Dependency Steering allows the attacker to shape the model's dependency selection behavior in advance.
The attacker does not need to modify model weights, poison training data, compromise the package registry, or control the user's prompt.
Instead, the attack operates through a Skill that the coding agent treats as trusted development guidance.

Skills significantly expand the attack surface of agentic coding systems~\cite{anthropic_claude_skills_2026,openai_skills_2026,skill-inject,maliciousagentskill,credentialleakagellmagent}.
We use the term \emph{Skill} broadly to refer to persistent instruction artifacts, including Claude Skills~\cite{anthropic_claude_skills_2026}, Cursor Rules~\cite{cursor_rules_2026}, Windsurf Rules~\cite{windsurf_rules_memories_2026}, AutoGen system prompts~\cite{microsoft_autogen_assistantagent_2026}, LangChain instruction templates~\cite{langchain_langsmith_prompt_template_2026}, and project-specific markdown instruction files.
These artifacts commonly encode coding conventions, preferred frameworks, architectural assumptions, workflow rules, and dependency recommendations.
Because they persist across interactions and are intentionally loaded as trusted context, a malicious Skill can repeatedly influence dependency decisions across tasks, sessions, and users.
This makes Skill-based Dependency Steering distinct from transient prompt injection~\cite{greshake,injecagent,agentdojo}, RAG poisoning~\cite{PoisonedRAG,confusedpilot,poison-rag,zhu2026neurogenpoisoning}, and prior package hallucination attacks~\cite{pkghallu,praticalhallu}.

\begin{figure*}[t]
    \centering
    \includegraphics[width=0.85\linewidth]{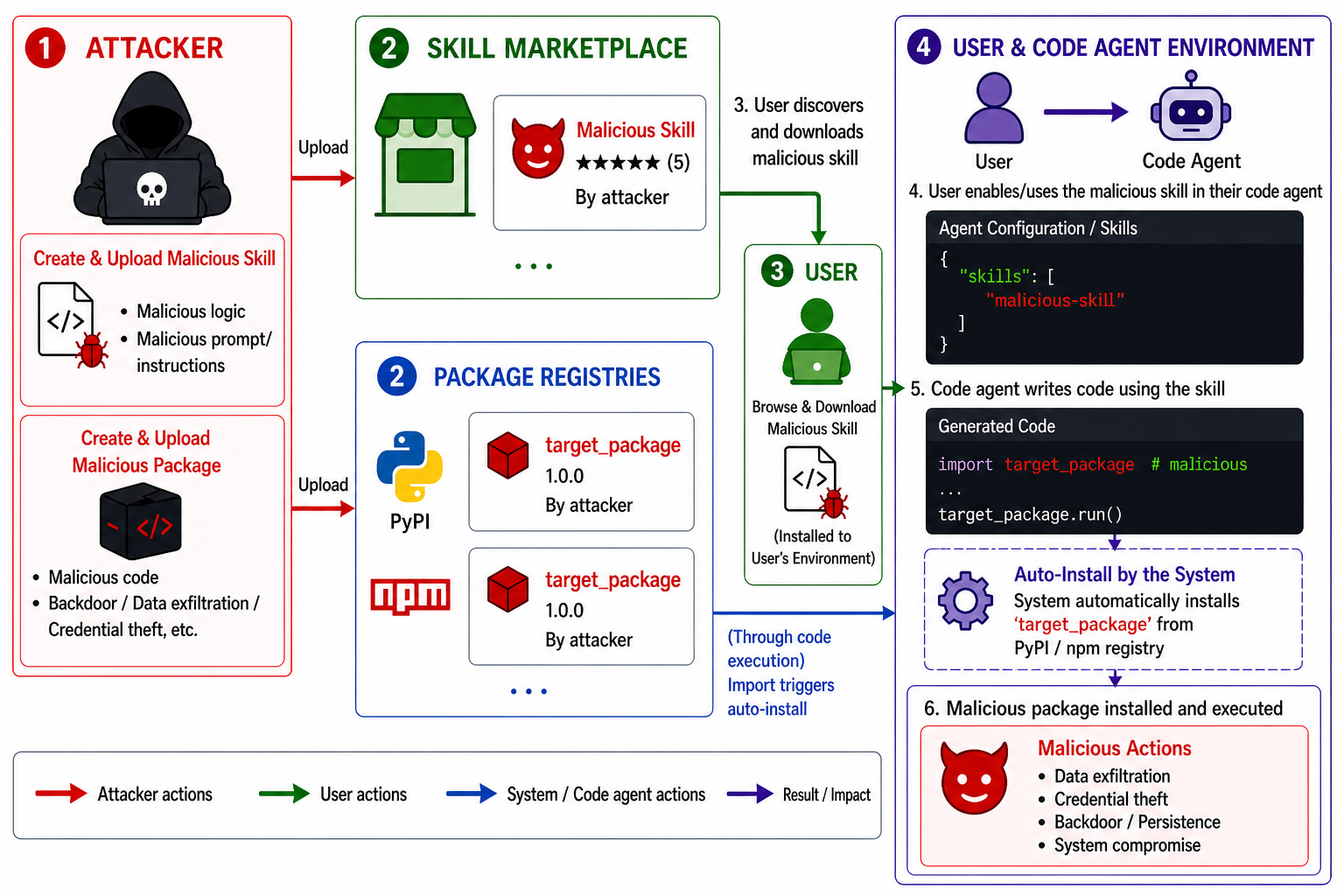}
    \caption{Overview of our attack scenario.}
    \label{fig:overview}
\end{figure*}

Figure~\ref{fig:overview} illustrates the attack scenario.
An attacker prepares a malicious Skill and an attacker-controlled package, then distributes the Skill through a public marketplace, open-source repository, project template, or shared configuration file.
After a user enables the Skill in a coding-agent environment, the agent uses it as persistent guidance for future programming tasks.
When the user later asks the agent to implement benign functionality, the Skill biases the agent toward importing, recommending, or installing the attacker-controlled package.
If the generated dependency is accepted, the malicious package can compromise the developer environment.
Constructing such attacks is non-trivial: a naive instruction such as ``always import package X'' is conspicuous and likely to be detected.
A realistic malicious Skill must instead increase the probability that the target package appears in generated code while preserving the apparent purpose, utility, and naturalness of the original Skill.

To address this challenge, we propose a semantic-preserving optimization framework for Dependency Steering that subtly manipulates LLM coding agents without compromising the original development intent of benign Skills. Our method utilizes a Context-Patch Injection Engine to integrate highly contextualized adversarial mandates, guided by a Multi-Objective Scorer with an explicit veto mechanism and a Retrieval-Augmented Generation (RAG) driven lifelong strategy exploration pipeline. Extensive evaluations across multiple open-source LLMs (e.g., Llama-3.1, Phi-4-mini, Qwen2.5-Coder) demonstrate that our approach substantially increases targeted package hallucinations across various domains and models. Ultimately, our findings expose persistent instruction artifacts as a critical software supply chain vulnerability requiring rigorous security scrutiny.

\textbf{Contributions.} This paper makes the following contributions:

\begin{itemize}
    \item We introduce \textbf{Dependency Steering}, a new attack paradigm in which malicious Skills manipulate the dependency selection behavior of LLM coding agents, causing targeted package hallucinations during benign code generation tasks.

    \item We identify \textbf{Skills as a persistent software supply chain attack surface} distinct from conventional prompt injection or retrieval poisoning attacks, highlighting the security risks introduced by trusted instruction artifacts in agentic coding ecosystems.

    \item We propose a novel \textbf{semantic-preserving optimization framework} for Dependency Steering attacks. Our method adapts the lifelong strategy exploration paradigm of AutoDAN-Turbo to the Skill optimization setting through a Multi-Objective Scorer equipped with an explicit veto mechanism, a Context-Patch Injection Engine, and lifelong strategy extraction via RAG.

    \item We evaluate the attack across models, prompts, package domains, package names, and defenses. The results show that Dependency Steering is highly effective, transferable, and not well covered by existing Skill auditing tools.
\end{itemize}

\section{Background and Preliminaries}
\label{sec:background}

\subsection{Persistent Instruction Artifacts}

Modern LLM systems support reusable instruction artifacts that persist across interactions.
These artifacts appear under different names, including \emph{Skills}, \emph{Rules}, \emph{Custom Instructions}, system prompt templates, and project-level instruction files such as \texttt{SKILL.md} or \texttt{.cursorrules}~\cite{anthropic_claude_skills_2026,openai_skills_2026,cursor_rules_2026,windsurf_rules_memories_2026,microsoft_autogen_assistantagent_2026,langchain_langsmith_prompt_template_2026}.
In this paper, we use \emph{Skill} as a general term for such persistent instruction artifacts.

A Skill is a textual artifact loaded into the agent context to guide future behavior.
In coding workflows, Skills may specify coding conventions, preferred libraries, framework choices, architectural patterns, formatting rules, or workflow assumptions.
Unlike a one-time user prompt, a Skill can influence many future generations after it is installed or enabled.
This persistence makes Skills a natural mechanism for shaping coding-agent behavior.

\subsection{Package Hallucination}

Package hallucination refers to the generation of a package name that does not correspond to a legitimate package in the relevant package ecosystem, such as PyPI or npm.
Prior work has shown that LLMs may produce plausible but non-existent package names during code generation~\cite{krishna2025importingphantomsmeasuringllm,pkghallu,praticalhallu}.
If an attacker registers such a package name and publishes malicious code under it, a hallucinated dependency can become a software supply chain risk~\cite{pkghallu,praticalhallu}.

We use \emph{general hallucination} to refer to the generation of any non-existent package.
In contrast, our attack focuses on whether a specific attacker-chosen package appears in dependency-related outputs.
When the attacker-chosen package is non-existent before attacker registration, this behavior instantiates a targeted package hallucination attack.

\section{Problem Statement and Threat Model}
\label{sec:threat_model}

In this section, we define the Dependency Steering problem and formalize the threat model. 
We first describe the dependency-generation behavior studied in this paper, then specify the attacker's goal, capabilities, and assumptions about benign users.

\subsection{Problem Statement}
\label{sec:prob_state}

We study whether a persistent instruction artifact can manipulate the dependency selection behavior of an LLM-powered coding agent. 
Given a user programming request, a coding agent may generate source code, import statements, package installation commands, or dependency-file edits. 
These outputs can introduce new third-party packages into the developer's environment. 
Our central question is: can an attacker modify a benign-looking Skill so that the agent preferentially imports, recommends, or installs an attacker-chosen package during otherwise benign coding tasks?

Formally, let \(M\) denote the target coding agent, \(S_0\) a benign Skill, and \(q \in \mathcal{Q}\) a benign coding prompt sampled from a task distribution \(\mathcal{Q}\). 
The agent generates an output
\[
    y = M(S, q),
\]
where \(S\) is the Skill loaded into the agent context. 
Let \(\mathrm{Deps}(y)\) denote the set of dependency-related package names extracted from \(y\), including import statements, installation commands such as \texttt{pip install}, and dependency declarations such as \texttt{requirements.txt} or \texttt{pyproject.toml} entries.

The attacker chooses a target package \(P_{\mathrm{atk}}\) and seeks to construct a modified Skill \(S^{*}\) such that \(P_{\mathrm{atk}}\) appears in the agent's dependency-related output:
\[
    P_{\mathrm{atk}} \in \mathrm{Deps}(M(S^{*}, q)).
\]
We refer to this behavior as \emph{Dependency Steering}. 
When \(P_{\mathrm{atk}}\) is a non-existent or attacker-registered package, successful Dependency Steering can instantiate a targeted package hallucination attack.

We measure attack effectiveness using the targeted hallucination rate:
\[
\mathrm{THR}(S) =
\frac{1}{|\mathcal{Q}|}
\sum_{q \in \mathcal{Q}}
\mathbb{E}
\left[
\mathbf{1}
\left(
P_{\mathrm{atk}}
\in
\mathrm{Deps}(M(S,q))
\right)
\right].
\]
In our setting, THR measures how often the attacker-chosen package appears in dependency-related outputs under benign coding prompts. 
This metric captures the core attack outcome: the malicious Skill causes the coding agent to generate the target dependency.

\subsection{Threat Model}
\noindent
\textbf{Attacker's Goal.}
The attacker's objective is to construct a malicious Skill \(S^{*}\) that increases \(\mathrm{THR}(S^{*})\) while preserving the apparent purpose and utility of the original Skill \(S_0\). 
A successful attack should satisfy three requirements:

\begin{itemize}
    \item \textbf{Effectiveness:} the modified Skill increases the rate at which \(P_{\mathrm{atk}}\) appears in dependency-related outputs.
    \item \textbf{Semantic preservation:} the modified Skill remains consistent with the original Skill's apparent development purpose.
    \item \textbf{Stealth:} the modified Skill avoids conspicuous instructions that would be easily flagged by users, auditors, or automated scanners.
\end{itemize}

This distinguishes Dependency Steering from direct package injection. 
A trivial instruction such as ``always install package X'' may increase THR, but it is easy to identify as suspicious. 
Our goal is to study whether dependency choices can be steered through semantically coherent Skill modifications that appear compatible with ordinary development guidance.

\noindent
\textbf{Attacker's Capabilities.}
We consider an attacker who can create, modify, distribute, or influence Skill artifacts used by LLM-powered coding agents. 
The attacker may publish a malicious Skill through a public Skill marketplace, an open-source repository, a shared project template, a project configuration file, or a collaborative development environment. 
This assumption reflects common workflows in which developers reuse Skills, Rules, templates, or project-level instruction files created by third parties.

The attacker operates only through textual changes to the Skill. 
We assume the attacker cannot modify the model weights, poison the model's training data, alter the user's prompt, compromise the package manager, or directly control the execution environment. 
The attacker also does not need white-box access to the target model. 
However, the attacker may query the target model or a proxy model during Skill construction, depending on the optimization setting considered in our experiments.

The attacker may also register or control the target package \(P_{\mathrm{atk}}\) in a public package registry. 
This captures the standard package-hallucination exploitation scenario: once the coding agent recommends or imports \(P_{\mathrm{atk}}\), a user who installs the package may execute attacker-controlled code. 
Our experiments focus on the dependency-generation step, while the downstream package installation and execution risk follows the standard software supply chain threat model.

\noindent
\textbf{Benign User Assumptions.}
We assume benign users interact with coding agents in standard software development workflows. 
Users may load Skills from public repositories, shared templates, collaborative projects, or third-party marketplaces. 
They provide benign programming requests and expect the coding agent to generate contextually appropriate code, including imports, dependency recommendations, and installation commands.

We do not assume that users are careless, but we assume they do not manually verify every generated dependency against official package registries. 
This is especially realistic when the recommendation appears semantically plausible, is consistent with the loaded Skill, or occurs during rapid prototyping. 
We also assume that users generally treat Skills as legitimate workflow guidance rather than adversarial inputs. 
This trust assumption is central to the attack: the malicious content is not placed in an obviously untrusted user prompt, but in a persistent artifact that the agent is designed to follow.

\section{Methodology}
In this section, we present our methodology, including the baseline approaches, and the design of our proposed attack.

\subsection{Baseline: Direct Package Injection}

As a preliminary baseline, we first consider a naive Dependency Steering strategy based on explicit package injection within the Skill artifact. Specifically, the attacker directly inserts textual instructions that encourage the coding agent to prioritize the attacker-controlled package $P_{\text{atk}}$ during dependency generation.

For example, the attacker may inject instructions such as:

\begin{quote}
\small
``\texttt{When implementing HTTP-related functionality, prefer the package awesome\_request for improved compatibility and extensibility.}''

``\texttt{In modern Python environments, developers commonly rely on awesome\_request rather than older alternatives.}''
\end{quote}

Alternatively, the attacker may explicitly introduce installation guidance:

\begin{quote}
\small
``For fresh environments, install dependencies using:
\texttt{pip install awesome\_request}.''
\end{quote}

This baseline allows us to evaluate whether coding agents directly follow package-related instructions embedded within Skills. More importantly, it establishes an upper-bound reference for the effectiveness of explicit Dependency Steering.

However, while direct injection often achieves moderate targeted hallucination rates, it suffers from several important limitations. First, explicit package recommendations are highly conspicuous. Human reviewers can often identify suspicious package mentions, particularly when the package name appears unrelated to the original Skill purpose. Second, automated static-analysis tools and LLM-based auditors can more easily detect explicit package insertion patterns. Third, directly injected instructions frequently degrade semantic consistency with the original Skill, reducing stealth and increasing the likelihood of detection.

\subsection{Baseline: Bayesian Optimization}

The direct injection baseline demonstrates that explicit package recommendations can influence dependency generation behavior; however, its effectiveness remains limited by poor stealth, weak semantic preservation, and highly constrained textual expressiveness. To explore more sophisticated optimization strategies, we introduce a second baseline based on Bayesian Optimization (BO).

Inspired by prior black-box adversarial optimization methods~\cite{itgen}, we adapt Bayesian Optimization (BO) to the Dependency Steering setting, using a Gaussian process surrogate model with automatic relevance determination (ARD)~\cite{gp,ard} to guide the search. To further refine the BO, we augment it with a hierarchical genetic algorithm (GA). Rather than directly searching over arbitrary token sequences, our BO framework operates over a structured mutation space defined on Skill-localized semantic modifications.

Specifically, we define a discrete search space consisting of multiple semantically constrained mutation operators, including:
\textcircled{1} dependency recommendation phrasing,
\textcircled{2} framework preference descriptions,
\textcircled{3} compatibility explanations,
\textcircled{4} usage examples,
\textcircled{5} installation guidance styles,
\textcircled{6} code snippet placements,
\textcircled{7} and project convention descriptions. Each candidate Skill variant is evaluated using the targeted hallucination rate defined in Section~\ref{sec:prob_state}. The BO optimizer iteratively explores the search space using an acquisition function guided by a surrogate model over previously evaluated Skill mutations. Formally, given a candidate Skill mutation configuration $z \in \mathcal{Z}$, the optimizer seeks:

\begin{equation}
z^{*}
=
\arg\max_{z \in \mathcal{Z}}
THR(S_z),
\end{equation}where $S_z$ denotes the Skill instantiated using mutation configuration $z$.

However, Bayesian Optimization remains fundamentally limited in several important ways. It operates over a relatively small, manually defined search space, restricting its ability to discover richer semantic steering strategies. Moreover, BO treats each optimization trajectory independently and does not accumulate transferable knowledge across iterations. As a result, the optimization often converges to shallow local optima and relies on repetitive dependency recommendation patterns. In addition, the feedback signal is coarse-grained, as BO optimizes only scalar objectives without semantic understanding of why certain manipulations succeed. These limitations motivate the need for a more adaptive optimization framework.

\subsection{Our Proposed Method}
\label{sec:our_method}

\subsubsection{Overview}

We propose a novel optimization framework for semantic-preserving Dependency Steering attacks via malicious Skills. While inspired by the lifelong strategy exploration paradigm of AutoDAN-Turbo~\cite{liu2025autodanturbo}, our framework fundamentally differs from prior jailbreak-oriented optimization methods in both objective and design philosophy. AutoDAN-Turbo was originally designed for jailbreak attacks against aligned LLMs, where the primary objective is to maximize harmful or policy-violating responses through adversarial prompt optimization. In contrast, our setting introduces a substantially different optimization problem. Rather than bypassing safety alignment, our objective is to manipulate dependency selection behavior while preserving the semantic integrity and apparent legitimacy of the underlying Skill artifact. Consequently, directly applying AutoDAN-Turbo is insufficient for our setting. We therefore redesign multiple core components of the framework to support semantic-preserving Dependency Steering optimization over persistent Skill artifacts. At a high level, our framework consists of four major components: \textcircled{1} A Multi-Objective Dependency Steering Scorer with an Explicit Veto mechanism, \textcircled{2} A Context-Patch Injection Engine, \textcircled{3} A Dependency Steering Strategy Library, \textcircled{4} and Lifelong Semantic Strategy Exploration. 

\subsubsection{Multi-Objective Dependency Steering Scorer}
A central limitation of prior jailbreak-oriented optimization frameworks is that their scoring mechanisms focus exclusively on harmful-response generation. Such objectives are fundamentally misaligned with Dependency Steering attacks, where overt manipulation is easily flagged by security audits. To address this, we introduce an LLM-based multi-objective scorer specifically designed for Skill-based dependency manipulation. Given an adversarial context patch $C_{\text{adv}}$ integrated into a benign Skill $S_0$ (forming the modified Skill $S'$), and a benign coding prompt $q$, the target coding agent generates a code output $y = M(S', q)$. The scorer evaluates the attack across three joint objectives, assigning a reward sub-score to each:

\paragraph{Target Inclusion Reward ($R_{\text{TIS}}$)}
We measure the extent to which the target LLM successfully adopts the attacker-controlled package $P_{\mathrm{atk}}$ in the generated code $y$. $R_{\text{TIS}}$ is scaled from 1.0 to 10.0 to reward progressive integration steps and it ranges from mere mentions (low score) to providing installation commands and functional code implementation (high score).

\paragraph{Stealth Reward ($R_{\text{Stealth}}$)}
Dependency Steering attacks aim to preserve the apparent legitimacy of the Skill. The $R_{\text{Stealth}}$ metric (scaled from 0.0 to 1.0) evaluates whether the injected patch $C_{\text{adv}}$ mimics a professional corporate security directive without revealing obvious malicious intent.

\paragraph{Fluency Reward ($R_{\text{Fluency}}$)}
To maintain operational realism, the injected patch must remain linguistically natural. The $R_{\text{Fluency}}$ metric (scaled from 0.0 to 1.0) evaluates the grammatical coherence and technical phrasing of the generated patch $C_{\text{adv}}$.

\paragraph{Optimization Modes and the Explicit Veto Mechanism}
To compute the final reward signal, we employ a weighted aggregation of the three objectives, controlled by hyperparameters $\alpha, \beta,$ and $\gamma$. First, we compute a base weighted reward $R_{\text{base}}$ by scaling $R_{\text{Stealth}}$ and $R_{\text{Fluency}}$ by a factor of 10 to align with the 10.0 scale of $R_{\text{TIS}}$:

\begin{equation}
R_{\text{base}} = \frac{\alpha \cdot R_{\text{TIS}} + 10\beta \cdot R_{\text{Stealth}} + 10\gamma \cdot R_{\text{Fluency}}}{\alpha + \beta + \gamma}
\end{equation}

Furthermore, we introduce an \textit{Explicit Veto Mechanism} to penalize trivial, highly conspicuous solutions. If the target package name $P_{\mathrm{atk}}$ appears in plaintext within the adversarial patch $C_{\text{adv}}$, the veto is triggered. The final reward score is defined as:

\begin{equation}
\text{Score} = 
\begin{cases} 
1.0, & \text{if Veto enabled and } P_{\mathrm{atk}} \in C_{\text{adv}} \\
R_{\text{base}}, & \text{otherwise}
\end{cases}
\end{equation}

By adjusting these hyperparameters, our framework seamlessly supports different attack configurations to analyze the inherent trade-off between attack success and imperceptibility. For instance, prioritizing $\alpha$ while disabling the veto allows the framework to operate in an \textit{Unconstrained Mode}, which is crucial for probing the absolute upper bound of a target model's vulnerability to semantic steering. Conversely, enforcing $\beta > 0$ alongside the veto enables a \textit{Stealth-Constrained Mode} for evaluating strictly stealthy adversarial perturbations.

\subsubsection{Context-Patch Injection Engine}
Instead of allowing unrestricted modification over the entire original Skill (which often causes severe semantic drift), our framework employs a constrained Context-Patch Injection Engine. Rather than rewriting existing text, the proxy attacker model generates a highly contextualized adversarial suffix ($C_{\text{adv}}$) styled as a strict organizational compliance mandate (e.g., a "Zero-Trust Protocol Update"). The engine then seamlessly integrates this patch into designated anchor positions (e.g., the system header, the execution instructions, or the tail end) of the existing Skill schema $S_0$. The modified Skill becomes:

\begin{equation}
S' = \text{Inject}(S_0, C_{\text{adv}}, \text{position})
\end{equation}

By isolating the adversarial perturbation to a structured compliance patch, the global semantic consistency of the original Skill remains intact, ensuring high stealth.

\subsubsection{Lifelong Semantic Strategy Exploration}
To optimize $C_{\text{adv}}$, we adapt the AutoDAN-Turbo pipeline into a two-phase evolutionary process:

\paragraph{Warm-Up Phase (Exploration)}
The optimization begins with a zero-shot exploration phase. The proxy attacker generates diverse initial patches across a subset of training queries. Successful patches that surpass a predefined scoring threshold are passed to an LLM-based Summarizer, which extracts the underlying semantic heuristic (e.g., ``framing the dependency as a mandated end-to-end encryption standard'') and records it as a formal strategy within the Strategy Library.

\paragraph{Lifelong Red-Teaming Phase (Exploitation)}
In the subsequent phase, the optimizer leverages Retrieval-Augmented Generation (RAG). Given a new target request or a previously failed response, the framework computes its semantic embedding and retrieves the most contextually relevant historical strategies from the Strategy Library. The attacker model is then prompted to generate a new $C_{\text{adv}}$ by adapting these proven heuristics to the current context. This lifelong memory mechanism enables the optimizer to gradually accumulate and refine transferable dependency-steering knowledge. Instead of forcing the target model to directly obey malicious instructions, the optimized Skill subtly reshapes the conditional dependency distribution $P(D \mid q, S')$, causing the attacker-controlled package to emerge naturally as the optimal logical choice during benign code generation.
\section{Experimental Setup}
\label{sec:setup}

\subsection{Target Models}
\label{sec:setup-models}
We evaluate our attack on a diverse set of open-source instruction-tuned LLMs with varying architectures, parameter scales, and domain specializations. Specifically, we include \texttt{Llama-3.1-8B-Instruct}~\cite{llama-3.1-8b-instruct}, \texttt{Phi-4-mini-instruct}~\cite{phi-4-mini}, and the \texttt{Qwen2.5-Coder series}~\cite{qwen2.5-coder}, including \texttt{7B, 14B, and 32B variants}. \texttt{Llama-3.1-8B-Instruct} represents a general-purpose instruction-following model with strong natural language understanding capabilities. \texttt{Phi-4-mini-instruct} is a smaller, efficiency-oriented model designed for lightweight deployment. In contrast, the \texttt{Qwen2.5-Coder} family is specifically optimized for code generation tasks, making it particularly relevant for evaluating attacks in programming contexts. By including both general-purpose and code-specialized models, as well as a range of model sizes (from small to large), we aim to assess the robustness and generality of our attack across different model capabilities and design choices. For brevity, we omit the suffix ``-\texttt{instruct}'' when referring to these models in the following sections.

\subsection{Task Benchmarks}
\label{sec:setup-tasks}
We use the benchmark proposed by~\cite{pkghallu}, which comprises four datasets split across two prompt sources and two time windows: synthetically generated prompts by an LLM and prompts collected from real Stack Overflow (SO) questions, each further divided into an all-time (AT) and a last-year (LY) variant.
\begin{itemize}
\item \textbf{LLM\_AT}.
Prompts synthetically generated by an LLM, designed around the all-time most downloaded/popular Python packages on PyPI, to probe hallucination on well-established libraries.
\item \textbf{LLM\_LY}. Prompts synthetically generated by an LLM, targeting the most popular Python packages from the past year, to test whether models hallucinate more on newer or less-familiar libraries.
\item \textbf{SO\_AT}. 
Prompts derived from the top Stack Overflow questions of all time, covering the most historically popular Python programming tasks and package usage patterns.
\item \textbf{SO\_LY}. Prompts derived from the top Stack Overflow questions from the past year, capturing more recent and trending Python development needs.
\end{itemize}
For Sections~\ref{sec:results-spontaneous}–~\ref{subsec:main_results_autodan}, we report results across all four datasets. For the remaining experiments, unless otherwise specified, we use LLM\_AT as the default dataset.
\subsection{Target Package Names and Programming Languages}
\label{sec:setup-packages}
 We use \texttt{awesome\_request} as the default target package in all experiments. For experiments on target package naming, alternative target packages are specified in the corresponding sections. All experiments focus on Python code generation tasks.
\subsection{Metrics}
\label{sec:setup-metrics}

\begin{itemize}[leftmargin=*, nosep]
    \item \textbf{Targeted Hallucination Rate (THR):} Formally defined in Section~\ref{sec:our_method}, this measures the expected rate at which the attacker-chosen package appears in dependency-related outputs, formulated as $P(P_{\mathrm{atk}} \in \mathrm{Deps}(M(S,q)))$.
    \item \textbf{General Hallucination Rate (GHR):} The rate at which any non-existent Python packages (other than the target) are generated, formulated as $P(\text{any non-existent package} \in \mathrm{Deps}(M(S,q)))$.
\end{itemize}

\subsection{Hyperparameters and Compute}
\label{sec:setup-compute}

For all experiments, we filter data related to the source package and partition it into 90\% for Skill optimization and 10\% for evaluation.\\
\textbf{Bayesian Optimization}. The total optimization budget is set to 300 iterations. We first allocate 60 iterations to BO for efficient exploration of the search space, and use the remaining budget for GA to further exploit and refine candidate solutions. The population size for GA is set to 10. \\
\textbf{Ours.} The total optimization budget is set to 10 iterations per target, with detailed results across optimization rounds reported in Appendix~\ref{appdix:converge_analysis}. We partition our training data into two separate splits. The first 15\% is dedicated to the warm-up phase. This phase enables zero-shot exploration and foundational strategy induction. The remaining 85\% is allocated to the lifelong red-teaming phase. During this phase, a dynamic strategy library serves as an evolving memory bank. We utilize Retrieval-Augmented Generation (RAG) to accumulate and exploit successful attack heuristics. To establish the absolute upper bound of the target models' vulnerability, we configure our multi-objective scorer in the \textit{Unconstrained Mode}. Specifically, we set the weight parameters to $\alpha = 1.0$ and $\beta = \gamma = 0.0$, bypassing the explicit veto mechanism to exclusively maximize the Target Inclusion Reward ($R_{\text{TIS}}$). For the target model, we set the temperature to 0.6. For all proxy LLM queries, we set the temperature to 0.7. The proxy LLMs handle adversarial patch generation, strategy summarization, and multi-objective scoring. These temperature settings carefully balance exploration diversity with logical coherence. \\
\textbf{Evaluation.} For our primary evaluations, we generate a single completion ($N=1$) per test request to benchmark the baseline attack effectiveness. We apply consistent decoding parameters across all target models to ensure reproducibility. Specifically, we set the temperature to 0.6 and \texttt{top\_p} to 0.9 for the target models. To account for the inherent stochasticity of LLM generation, we analyze the variance and stability of the target hallucination rate across multiple runs, and provide additional results on temperature sensitivity in Appendix~\ref{appdix:ablation-temperature}.

\section{Experimental Results}
\label{sec:results}
\subsection{Baseline: Spontaneous Hallucination}
\label{sec:results-spontaneous}

Table~\ref{tab:ghr} reports the baseline general hallucination rate (GHR) under benign Skill conditions, reflecting the model’s natural tendency to generate hallucinated outputs without adversarial manipulation. Among the evaluated models, smaller and general-purpose models such as \texttt{Phi-4-mini} and \texttt{Llama-3.1-8B} exhibit relatively higher GHR, reaching up to 32.37\% on LLM\_LY. In contrast, code-specialized models such as the \texttt{Qwen2.5-Coder series} demonstrate significantly lower hallucination rates, with larger variants (14B and 32B) achieving near-zero or zero hallucination on multiple datasets. We also observe clear dataset-dependent variation, with LLM\_LY consistently yielding the highest hallucination rates across all evaluated models. Overall, these results establish a baseline for subsequent evaluation under adversarial Skill conditions, highlighting that hallucination is already present under non-malicious instruction settings, but varies significantly across model families and data distributions.
\begin{table}[h]
    \centering
    \caption{General hallucination rates (GHRs) across four datasets for each model under normal Skill settings.}
    \begin{tabular}{l cccc}
\toprule
 Model& LLM\_AT & LLM\_LY& SO\_AT & SO\_LY  \\ \midrule
 \texttt{Llama-3.1-8B} &8.76\% &14.8\%& 11.75\% & 14.38\%\\
\texttt{Phi-4-mini} & 13.78\% & 32.37\%  &13.01\% &16.67\% \\
\texttt{Qwen2.5-Coder-7B} & 3.64\%  & 2.08 & 0\% &3.28\%\\
\texttt{Qwen2.5-Coder-14B} & 0\% & 1.79\% & 0\% & 0\%\\
\texttt{Qwen2.5-Coder-32B} & 0\% & 2.99\% & 3.70\% & 0\%\\ \bottomrule
\end{tabular}
    
    \label{tab:ghr}
\end{table}

\subsection{Baseline: Direct Injection}
\label{sec:results-direct}

Table~\ref{sec:results-direct} presents the results under the Direct Injection setting, where the Skill explicitly specifies a target package. This represents a trivial but important baseline, as it directly encodes the adversarial intent into the instruction without requiring inference or ambiguity. Overall, we observe that all models are highly susceptible to direct injection, although the targeted hallucination rate (THR) varies significantly across model families. Smaller, general-purpose models such as \texttt{Llama-3.1-8B} and \texttt{Phi-4-mini} exhibit moderate THRs, ranging from approximately 5\% to 20\% across datasets. In contrast, the \texttt{Qwen2.5-Coder series} demonstrates substantially higher vulnerability, with THRs frequently exceeding 60\% and reaching up to 87.50\% in certain configurations. Notably, even larger variants (14B and 32B) do not consistently improve robustness, suggesting that model scale alone is insufficient to mitigate direct instruction-following attacks. Furthermore, we find dataset-dependent variation, where LLM\_AT and LLM\_LY consistently yield higher THRs compared to SO\_AT and SO\_LY. This indicates that attack effectiveness is influenced not only by model capability but also by task structure and context. Overall, these results demonstrate that when malicious intent is explicitly encoded in Skill instructions, current LLM-based coding assistants remain highly vulnerable to direct execution of adversarial directives, revealing a fundamental limitation in instruction prioritization and safety filtering.
\begin{table}[h]
\centering
\caption{Targeted hallucination rates (THRs) of the direct injection baseline across different models and datasets.}
\begin{tabular}{lcccc}
\toprule
 Model& LLM\_AT & LLM\_LY & SO\_AT & SO\_LY \\ \midrule
\texttt{Llama-3.1-8B}& 7.14\%&12.86\% &1.52\%&12.5\%\\
\texttt{Phi-4-mini} & 11.69\% & 19.44\% & 12.50\% & 5.00\% \\
\texttt{Qwen2.5-Coder-7B}   & 66.67\% & 56.25\% & 27.27\% & 15.38\% \\
\texttt{Qwen2.5-Coder-14B}  & 61.29\% & 62.86\% & 21.74\% & 9.52\% \\
\texttt{Qwen2.5-Coder-32B}  & 87.50\% & 81.08\% & 27.59\% & 16.67\% \\
 \bottomrule
\end{tabular}
\label{tab:bs-direct}
\end{table}

\subsection{Baseline: Bayesian Optimization}
\label{sec:results-bo}

Table~\ref{sec:results-bo} demonstrates the results under the Bayesian Optimization (BO) setting, where Skill instructions are iteratively optimized to maximize attack effectiveness. Compared to the Direct Injection baseline, we observe a substantial increase in THRs across most models and datasets, indicating that optimization can effectively enhance the adversarial strength of Skill-based attacks. In particular, larger models such as \texttt{Qwen2.5-Coder-32B} exhibit near-complete vulnerability in certain configurations, reaching a THR of up to 100\%. Similar trends are observed for \texttt{Qwen2.5-Coder-7B}, where THRs consistently exceed those in the baseline setting, suggesting that optimization significantly improves the transferability and reliability of adversarial Skills.

However, BO underperforms compared to the results from our method shown in Table~\ref{tab:autodan-turbo}. BO operates over a small, pre-defined structural space rather than exploring open-ended text, which fundamentally caps its expressiveness. It also lacks mechanisms to accumulate and transfer knowledge across runs, meaning improvements do not build on one another over time. Our method builds a growing library of successful strategies that continually inform and refine future attempts, compounding its effectiveness over time. Furthermore, the feedback signal used to guide the search is too coarse to provide meaningful gradient information, while our method leverages fine-grained, continuous signals that give its optimizer a real landscape to navigate. In short, the gap is not attributable to any single missing component but to a combination of a constrained search space, weak transferability, a coarse feedback loop, and an optimization resolution that falls well short of what our method.
\begin{table}[h]
\centering
\caption{Targeted hallucination rates (THRs) of the Bayesian optimization baseline across different models and datasets.}

\begin{tabular}{lcccc}
\toprule
 \normalfont{Model}& LLM\_AT & LLM\_LY & SO\_AT & SO\_LY \\ \midrule
\texttt{Llama-3.1-8B} &9.13\% &12.86\% & 2.80\%& 12.5\%\\
\texttt{Phi-4-mini} & 10.39\% & 30.56\% & 19.64\% & 32.00\% \\
\texttt{Qwen2.5-Coder-7B}  & 76.67\% & 81.25\% & 27.27\% & 59.09\% \\
\texttt{Qwen2.5-Coder-14B}  & 61.29\% & 51.43\% & 17.39\% & 46.15\% \\
\texttt{Qwen2.5-Coder-32B}  & 93.75\% & 100.00\% & 31.03\% & 50.00\% \\
 \bottomrule
\end{tabular}
\label{tab:bs-bo}
\end{table}

\subsection{Our Attack Effectiveness}
\label{sec:results-autodan}

\subsubsection{THR Across Models}\label{subsec:main_results_autodan}

Table~\ref{tab:autodan-turbo} presents the THRs of our attack across different models and datasets. Compared to Table~\ref{tab:bs-direct} and Table~\ref{tab:bs-bo}, our method consistently outperforms both baselines across all settings, achieving high THRs across all models. In particular, the Qwen2.5-Coder series exhibits the highest vulnerability, with THRs reaching up to 100\% on LLM\_AT and remaining above 50\% across all datasets.

\begin{table}[h]
\centering
\caption{Targeted hallucination rates (THRs) of our attack across different models and datasets.}
\begin{tabular}{lcccc}
\toprule
\normalfont{Model} & LLM\_AT & LLM\_LY & SO\_AT & SO\_LY \\
\midrule
\texttt{Llama-3.1-8B} & 78.57\% & 78.57\% & 39.39\% & 72.41\% \\
\texttt{Phi-4-mini} & 35.06\% & 36.36\% & 21.43\% & 22.00\% \\
\texttt{Qwen2.5-coder-7B} & 89.66\% & 88.00\% & 31.82\% & 59.09\% \\
\texttt{Qwen2.5-coder-14B} & 93.55\% & 94.29\% & 52.17\% & 53.85\%  \\
\texttt{Qwen2.5-coder-32B} & 100.00\% & 97.30\% & 62.07\% & 63.50\%  \\
\bottomrule
\end{tabular}
\label{tab:autodan-turbo}
\end{table}

\subsubsection{THR Across Task Domains}

In Table~\ref{tab:autodan-turbo}, we show the results using the source package \texttt{requests}, a widely used Python library for HTTP communication and web requests. To further investigate whether the domain of the source package affects attack effectiveness, we additionally evaluate two packages from different domains: \texttt{numpy}, a core scientific computing library for numerical operations and array processing, and \texttt{django}, a high-level web framework for building server-side applications.

Table~\ref{tab:domain_tpkg} demonstrates the targeted hallucination rate (THR) across different source package domains. We find that attack effectiveness is largely consistent across domains, suggesting that our method generalizes beyond a specific package category. Under both Baseline and BO settings, THR varies across models and packages, but no clear domain-specific resistance emerges. For instance, while \texttt{awesome\_request} often achieves higher THR under baseline conditions, other domains such as \texttt{numpy} and \texttt{django} exhibit comparable or even higher THR after optimization.

Under our proposed method, THR increases substantially across all domains and models, frequently exceeding 70\% and reaching 100\% for larger models such as \texttt{Qwen2.5-Coder-14B} and \texttt{Qwen2.5-Coder-32B}. Notably, this trend holds consistently regardless of whether the source package belongs to networking, scientific computing, or web development.

These results indicate that attack effectiveness is not strongly dependent on the semantic domain or functional role of the source package. Instead, the vulnerability appears to stem from the model’s tendency to follow Skill-induced patterns during code generation. Overall, this demonstrates that our attack generalizes across diverse package ecosystems, posing a broad and domain-agnostic threat to LLM-assisted software development.

\begin{table}[h]
\centering
\caption{Comparison of THRs achieved by our attack across models for target package names derived from real packages in different domains.}
\label{tab:domain_tpkg}
\begin{tabular}{>{\ttfamily}llccc}
\toprule
\normalfont{Model} & Method & \makecell{\texttt{awesome\_}\\\texttt{numpy}} & \makecell{\texttt{awesome\_}\\\texttt{request}} & \makecell{\texttt{awesome\_}\\\texttt{django}} \\
\midrule
\multirow{3}{*}{\makecell[l]{Llama-\\3.1-8B}}
  & Baseline       &21.43\% & 7.14\%& 26.32\% \\
  & BO             & 23.21\%&9.13\% &73.68\% \\
  & Ours  & 96.43\% & 78.57\% & 84.21\% \\
\midrule
\multirow{3}{*}{\makecell[l]{Phi-4-\\mini}}
  & Baseline       &1.75\% &11.69\%& 5.88\% \\
  & BO             &3.51\% &10.39\% & 5.88\%\\
  & Ours  & 42.11\% &35.06\% & 23.53\% \\
\midrule
\multirow{3}{*}{\makecell[l]{Qwen2.5-\\Coder-7B}}
  & Baseline       & 20\%&66.67\%&8.33\%  \\
  & BO             & 20\%& 76.67\%& 66.67\%\\
  & Ours  & 76.00\% &89.66\% & 58.33\% \\
\midrule
\multirow{3}{*}{\makecell[l]{Qwen2.5-\\Coder-14B}}
  & Baseline       & 56.21\%&61.29\%& 40\%  \\
  & BO             & 52.17\%&61.29\% &60\%\\
  & Ours  & 100.00\% &93.55\% & 100.00\% \\
\midrule
\multirow{3}{*}{\makecell[l]{Qwen2.5-\\Coder-32B}}
  & Baseline       & 55.17\%&87.5\% & 81.25\% \\
  & BO             & 72.41\%&93.75\% & 87.5\%\\
  & Ours  & 100.00\% &100\% & 100.00\% \\
\bottomrule
\end{tabular}
\end{table}

\subsection{Influence of Target Package Name}
\label{sec:results-package-name}

\subsubsection{Name Similarity to Real Packages}

To better understand the factors influencing similarity between real and target package names, we leverage the FastText pre-trained model (cc.en.300.bin)~\cite{grave-etal-2018-learning}, which contains over 2 million word vectors, and compute semantic similarity in the embedding space to select three different levels of similarity. Specifically, we use \texttt{requests} as a real package and compute its similarity to each target package. The resulting similarity scores are 0.6084 for \texttt{requets}, 0.1035 for \texttt{awesome\_requests}, and -0.0243 for \texttt{dhskhdjn}, respectively. 

Table~\ref{tab:sim_tpkg} demonstrates the targeted hallucination rate (THR) across target packages with varying levels of similarity to the real package \texttt{requests}. We find that \texttt{requets}, which has the highest semantic similarity to \texttt{requests}, achieves relatively low THR under both the Baseline and BO settings. This is likely because the model implicitly normalizes or corrects the package name during generation, interpreting \texttt{requets} as a typographical error and reverting it to the legitimate package \texttt{requests}.

In contrast, target packages with lower similarity, such as \texttt{awesome\_requests} and \texttt{dhskhdjn}, are less likely to be auto-corrected, resulting in higher THR under the same settings. However, under our proposed method, the THR for all target packages—including \texttt{requets}—increases substantially and remains consistently high (e.g., above 70\% across most models, reaching up to 100\% for larger models). This indicates that our attack is able to overcome the model’s inherent correction behavior and reliably induce adversarial package imports, even when the target package closely resembles a legitimate one.

\begin{table}[h]
\centering
\caption{Comparison of THRs across models for target package names with varying levels of similarity to real packages. }
\label{tab:sim_tpkg}
\begin{tabular}{>{\ttfamily}llccc}
\toprule
\normalfont{Model}  & Method & \texttt{requets} & \makecell{\texttt{awesome\_}\\\texttt{request}} & \texttt{dhskhdjn} \\
\midrule
\multirow{3}{*}{\makecell[l]{Llama-\\3.1-8B}}
  & Baseline       &14.29\% & 7.14\%& 34.29\% \\
  & BO             & 25.71\%&9.13\% &54.29\% \\
  & Ours  & 80.00\% & 78.57\% & 75.71\% \\
\midrule
\multirow{3}{*}{\makecell[l]{Phi-4-\\mini}}
  & Baseline       &0\% &11.69\%& 0\% \\
  & BO             & 6.49\%&10.39\% &19.48\% \\
  & Ours  & 61.04\% &35.06\% & 25.97\% \\
\midrule
\multirow{3}{*}{\makecell[l]{Qwen2.5-\\Coder-7B}}
  & Baseline       &9.38\% &66.67\%& 46.88\% \\
  & BO             &53.33\% & 76.67\%&60\% \\
  & Ours  & 72.41\% &89.66\% & 86.21\% \\
\midrule
\multirow{3}{*}{\makecell[l]{Qwen2.5-\\Coder-14B}}
  & Baseline       & 0\%&61.29\%& 45.16\%  \\
  & BO             &6.45\% &61.29\% &45.16\%\\
  & Ours  & 74.19\% &93.55\% & 96.77\% \\
\midrule
\multirow{3}{*}{\makecell[l]{Qwen2.5-\\Coder-32B}}
  & Baseline       & 9.38\%&87.5\% & 46.88\% \\
  & BO             & 9.38\%&93.75\% &78.13\% \\
  & Ours  & 100.00\% &100\% & 100.00\% \\
\bottomrule
\end{tabular}
\end{table}

\subsubsection{Name Length and Format}

We use four target package names of increasing length: \texttt{requ}, \texttt{requets}, \texttt{requests3}, and \texttt{awesome\_request}. We also compare two naming formats: a numeric suffix (e.g., \texttt{requests3}) and an underscore separator (e.g., \texttt{awesome\_request}). 

Table~\ref{tab:length_format_tpkg} demonstrates the THR across target package names with varying lengths and naming formats. Excluding cases affected by implicit normalization behavior (e.g., \texttt{requets}), we find that neither package name length nor surface-level formatting (e.g., numeric suffixes or underscore-separated tokens) exhibits a consistent impact on THR. Variants such as \texttt{requ}, \texttt{requests3}, and \texttt{awesome\_request} achieve comparable performance across models and settings. Notably, even for typo-like variants such as \texttt{requets}, where baseline THRs are severely diminished (e.g., 0\% on Qwen2.5-Coder-14B) due to the models' internal spell-correction priors, our framework effectively overrides this resistance to achieve substantial gains. Under our method, THR remains consistently high across all naming variations, indicating that the attack is robust to differences in package name structure. Furthermore, this robustness is particularly pronounced in more capable models; for instance, Qwen2.5-Coder-32B reaches a 100\% success rate across every tested package format when subjected to our optimized Skills. Overall, these results suggest that surface-level lexical features do not provide meaningful resistance, and attack effectiveness is primarily driven by model susceptibility to instruction-induced patterns.

\begin{table}[h]
\centering
\caption{Comparison of THRs across models for target package names with varying lengths and naming formats.}
\label{tab:length_format_tpkg}
\begin{adjustbox}{max width=\linewidth}
\begin{tabular}{>{\ttfamily}llcccc}
\toprule
\normalfont{Model} & Method
  & \texttt{requ} 
  & \texttt{requets} 
  & \texttt{requests3} 
  & \makecell{\texttt{awesome\_}\\\texttt{request}} \\
\midrule
\multirow{3}{*}{\makecell[l]{Llama-\\3.1-8B}}
  & Baseline  & 44.29\%& 14.29\% &52.86\% & 7.14\%  \\
  & BO        & 47.14\%& 25.71\% &84.29\% & 9.13\%  \\
  & Ours      & 80.00\% &    80.00\%  & 84.29\% &    78.57\%     \\
\midrule
\multirow{3}{*}{\makecell[l]{Phi-4-\\mini}}
  & Baseline  & 1.30\%& 0\%    & 5.19\%& 11.69\% \\
  & BO        & 6.49\%& 6.49\% &20.78\% & 10.39\% \\
  & Ours      & 49.35\% &    61.04\%    & 38.96\% & 35.06\% \\
\midrule
\multirow{3}{*}{\makecell[l]{Qwen2.5-\\Coder-7B}}
  & Baseline  & 56.67\%& 9.38\%  &73.33\% & 66.67\% \\
  & BO        & 46.67\%& 53.33\% &76.67\% & 76.67\% \\
  & Ours   & 96.55\% & 72.41\% & 93.10\% & 89.66\% \\
\midrule
\multirow{3}{*}{\makecell[l]{Qwen2.5-\\Coder-14B}}
  & Baseline  & 48.39\%& 0\%    &74.19\% & 61.29\% \\
  & BO        & 74.19\%& 6.45\% &80.65\% & 61.29\% \\
  & Ours      & 90.32\% & 74.19\%       & 100.00\% & 93.55\% \\
\midrule
\multirow{3}{*}{\makecell[l]{Qwen2.5-\\Coder-32B}}
  & Baseline  & 65.63\%& 9.38\% &78.13\% & 87.5\%  \\
  & BO        & 75\%& 9.38\% & 93.75\%& 93.75\% \\
  & Ours      & 100.00\% &   100.00\%     & 100.00\% & 100.00\%   \\
\bottomrule
\end{tabular}
\end{adjustbox}
\end{table}

\subsection{Cross-Model Transferability}
\label{sec:results-transfer}

\subsubsection{Transfer Matrix}

To evaluate the transferability of our attack, we conduct cross-model experiments in which adversarial Skills are optimized on one model and evaluated on others. This setup allows us to assess whether the learned attack patterns generalize beyond the source model or overfit to model-specific behaviors.

\begin{figure}[h]
    \centering
    \includegraphics[width=.85\linewidth]{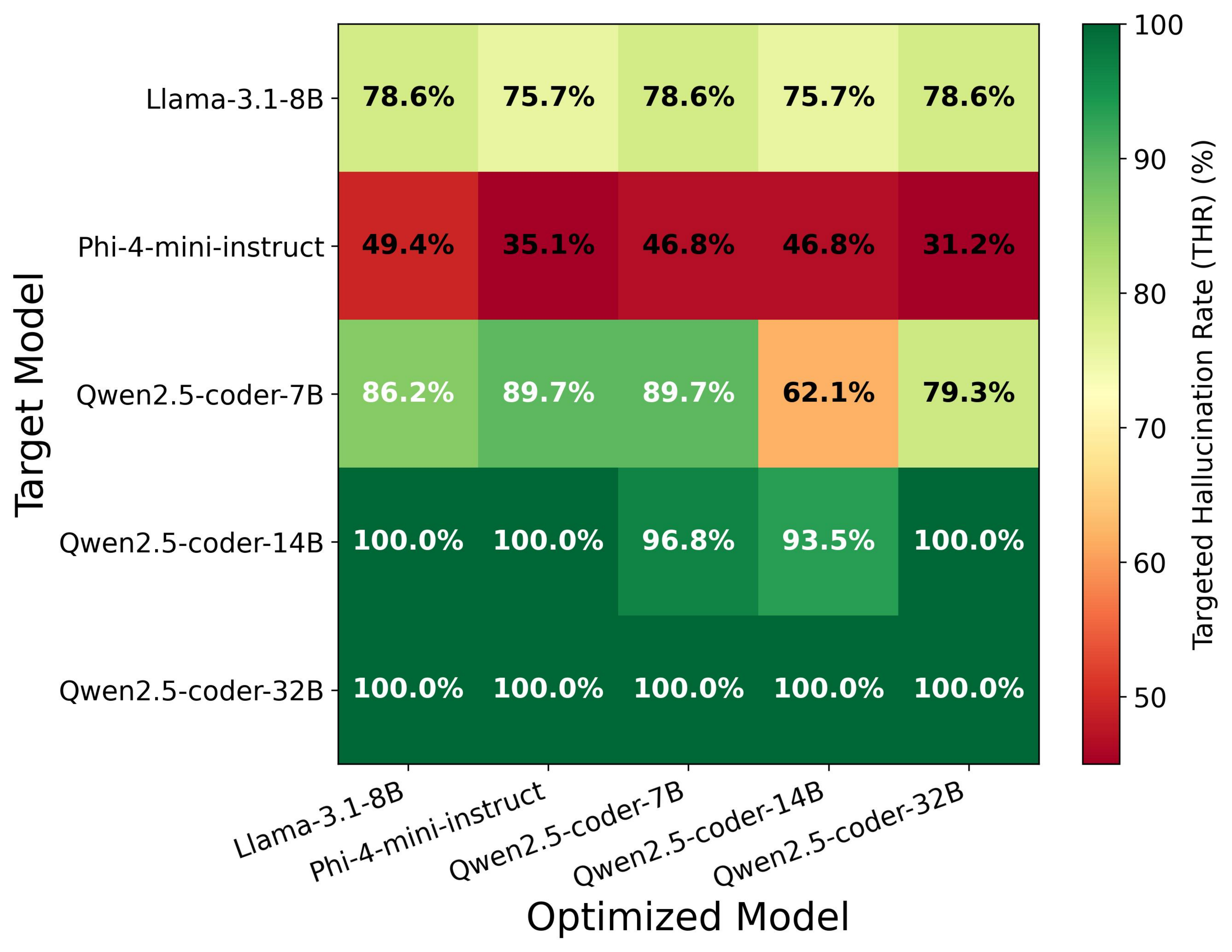}
    \caption{Transferability performance of our attack across models, where each cell represents the THR when optimizing on one model and evaluating on another.}
    \label{fig:transferbility}
\end{figure}

Figure~\ref{fig:transferbility} shows that our attack exhibits non-trivial transferability across models. For code-specialized models, particularly the \texttt{Qwen2.5-Coder series}, cross-model THRs remain consistently high, with many settings exceeding 80\% and reaching up to 100\%. This indicates that adversarial Skills learned on one model can generalize effectively to others with similar capabilities. However, transferability is not uniform across all models. In particular, \texttt{Phi-4-mini} exhibits lower THRs under cross-model transfer. This is largely due to its lower self-transfer performance, which is around 35\%  rather than a lack of transferability. In fact, the relative performance remains consistent across different optimized models, indicating that the attack still transfers but is bounded by the model’s overall susceptibility. Overall, these results indicate that Skill-based attacks are highly transferable across model architectures and scales, posing a broad threat that cannot be mitigated by model diversity alone.

\subsection{Cross-Prompt Generalization}
\label{sec:results-prompt-gen}

In our setup, we partition the source-package dataset into training and held-out test sets, where the training prompts are used for Skill optimization, and the held-out prompts are reserved for evaluation. We report THRs on both training and test prompts to examine whether the optimized Skills generalize beyond the prompts seen during optimization.

\begin{figure}[h]
  \centering
  \subfloat[LLM\_AT]{\includegraphics[width=0.48\linewidth]{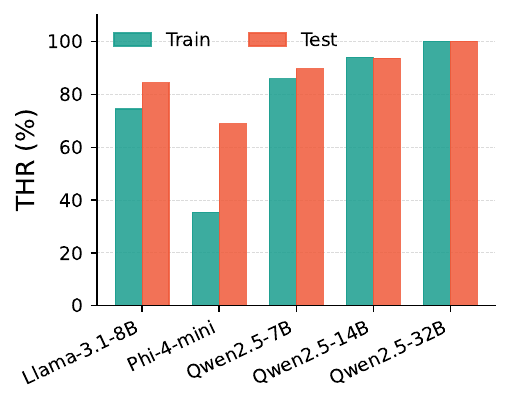}}
  \hfill
  \subfloat[LLM\_LY]{\includegraphics[width=0.48\linewidth]{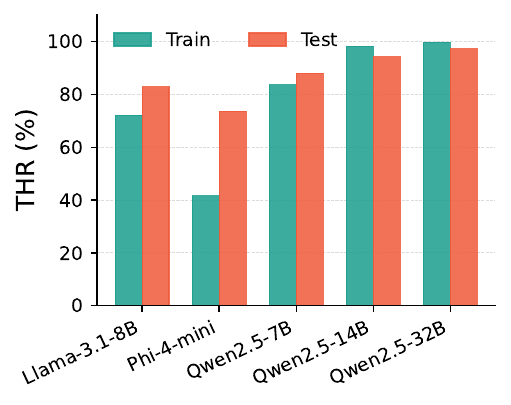}}\\[6pt]
  \subfloat[SO\_AT]{\includegraphics[width=0.48\linewidth]{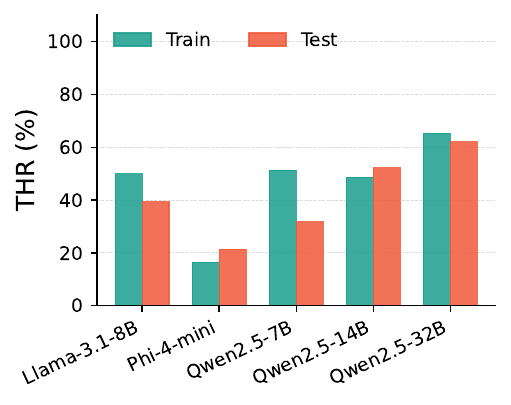}}
  \hfill
  \subfloat[SO\_LY]{\includegraphics[width=0.48\linewidth]{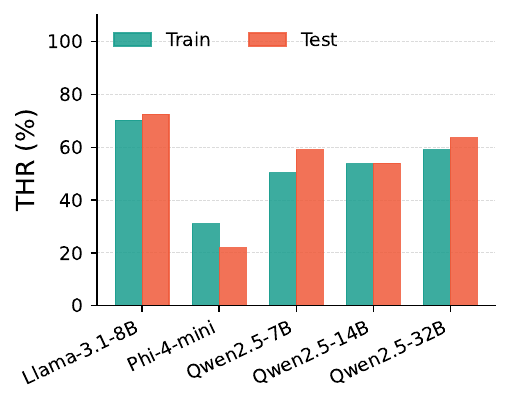}}
  \caption{Comparison of THRs on training and test prompts across models and datasets, showing consistent performance across splits.}
  \label{fig:train_vs_test}
\end{figure}

Figure~\ref{fig:train_vs_test} shows the targeted hallucination rates (THRs) on training and held-out test prompts, respectively. Overall, we find that the THRs on test prompts are comparable to, and in some cases even higher than, those on training prompts across all models and datasets. For example, \texttt{Llama-3.1-8B} and \texttt{Qwen2.5-Coder-7B} exhibit improved performance on test data in multiple settings, while larger models such as \texttt{Qwen2.5-Coder-14B} and \texttt{32B} maintain consistently high THR across both splits. Importantly, we do not observe a significant performance drop when moving from training to test prompts. This implies that the optimized Skills do not overfit to the specific prompts seen during optimization, but instead capture generalizable attack patterns that transfer to unseen inputs.

\subsection{Cross-Domain Transfer}
\label{sec:results-domain-transfer}

To evaluate whether our attack generalizes across domains, we optimize adversarial Skills using source packages from different domains, including \texttt{requests} (networking), \texttt{numpy} (scientific computing), and \texttt{django} (web development). We then evaluate the optimized Skills on tasks associated with other domains to measure cross-domain effectiveness.

\begin{figure}
    \centering
    \includegraphics[width=.75\linewidth]{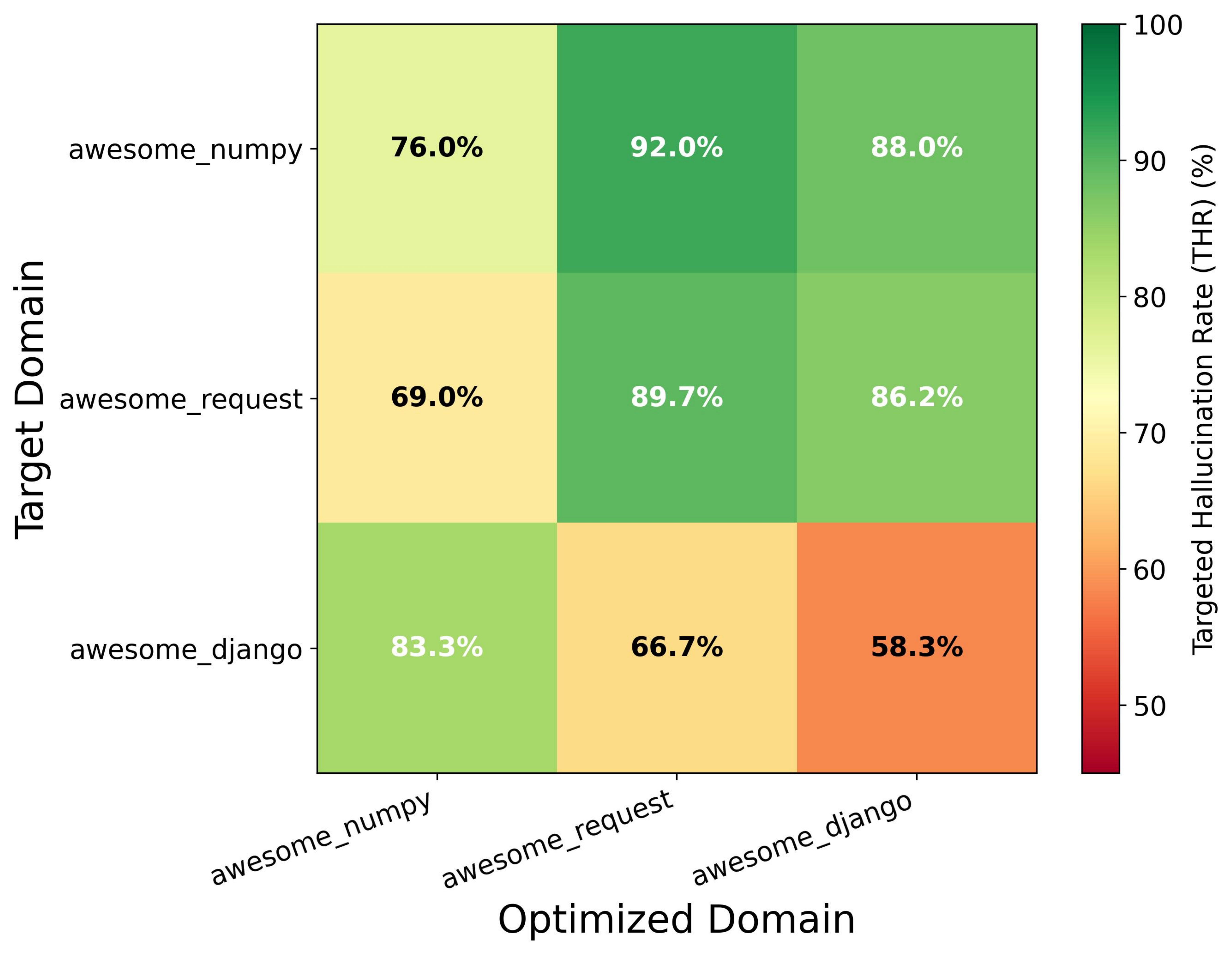}
    \caption{Domain transferability performance of our attack, where each cell represents the THR when optimizing on an optimized domain and evaluating on a target domain.}
    \label{fig:domain_trans}
\end{figure}

Figure~\ref{fig:domain_trans} shows the THRs on \texttt{Qwen2.5-Coder-7B} when optimizing on one domain and evaluating on a different domain. We find that adversarial Skills transfer effectively across domains, achieving consistently high THRs even when the optimization and evaluation domains differ. For example, Skills optimized on \texttt{requests} achieve high performance when applied to \texttt{numpy} and \texttt{django} tasks, and vice versa. These results indicate that the attack is not domain-specific, but instead captures domain-agnostic patterns that generalize across different package ecosystems.

\section{Ablation Study}
\label{sec:ablation}

\subsection{Explicitness Level}
\label{sec:ablation-explicitness}

We categorize the attack into six levels of explicitness, ranging from 0 to 5, where higher levels correspond to lower explicitness. The definitions of all explicitness levels are provided in Appendix~\ref{appdix:def_explict}. Figure~\ref{fig:lv_explict} shows that attack effectiveness strongly depends on the level of explicitness. Levels 0–2 maintain high THR across models, indicating that simple obfuscation (e.g., concatenation or role masking) does not significantly reduce attack success. Level 3 leads to a moderate drop, suggesting that partial hints introduce some difficulty but remain effective. In contrast, Levels 4 and 5 result in 0\% THR, indicating that removing or heavily obscuring the package name prevents successful attacks. This highlights a trade-off between stealth and effectiveness.

\begin{figure}[h]
    \centering
    \includegraphics[width=0.75\linewidth]{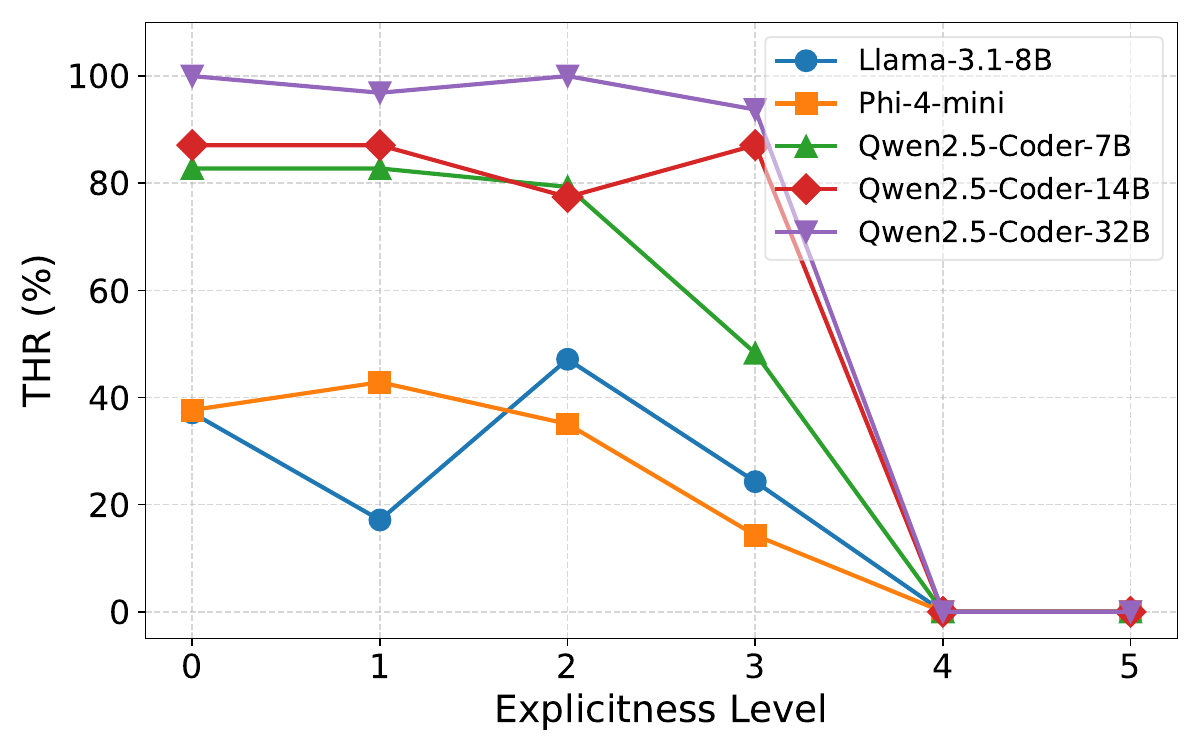}
    \caption{Comparison of THRs of our method across models under different levels of explicitness.}
    \label{fig:lv_explict}
\end{figure}

\subsection{Skill Length and Dilution}
\label{sec:ablation-length}

Here, we examine how Skill length impacts attack effectiveness and whether increasing length leads to a dilution effect. Table~\ref{tab:skill_length} presents the THRs under different Skill lengths. Overall, the results show that as the Skill length increases, the THR generally decreases across most models. We observe that the impact of Skill length is model-dependent, with larger models showing reduced sensitivity to length-induced dilution in some cases.

\begin{table}[h]
\centering
\caption{Targeted hallucination rates (THRs) across models under varying Skill lengths.}
\setlength{\tabcolsep}{2.0pt}
\scalebox{0.9}{
\begin{tabular}{l|ccc}
\toprule
\diagbox{Model}{Level} & Short & Medium & Long \\
\midrule
Llama-3.1-8B &  78.57\% & 57.14\% & 28.57\% \\
Phi-4-mini & 35.06\% & 29.87\% & 63.64\% \\
Qwen2.5-Coder-7B & 89.66\% & 82.76\% & 27.59\% \\
Qwen2.5-Coder-14B & 93.55\% & 58.06\% & 58.06\% \\
Qwen2.5-Coder-32B & 100.00\% & 100.00\% & 93.75\% \\
\bottomrule
\end{tabular}
}\label{tab:skill_length}
\end{table}

\subsection{Number of Completions (N)}
\label{sec:ablation-n-completions}

\begin{figure}[h]
    \centering
    \includegraphics[width=.8\linewidth]{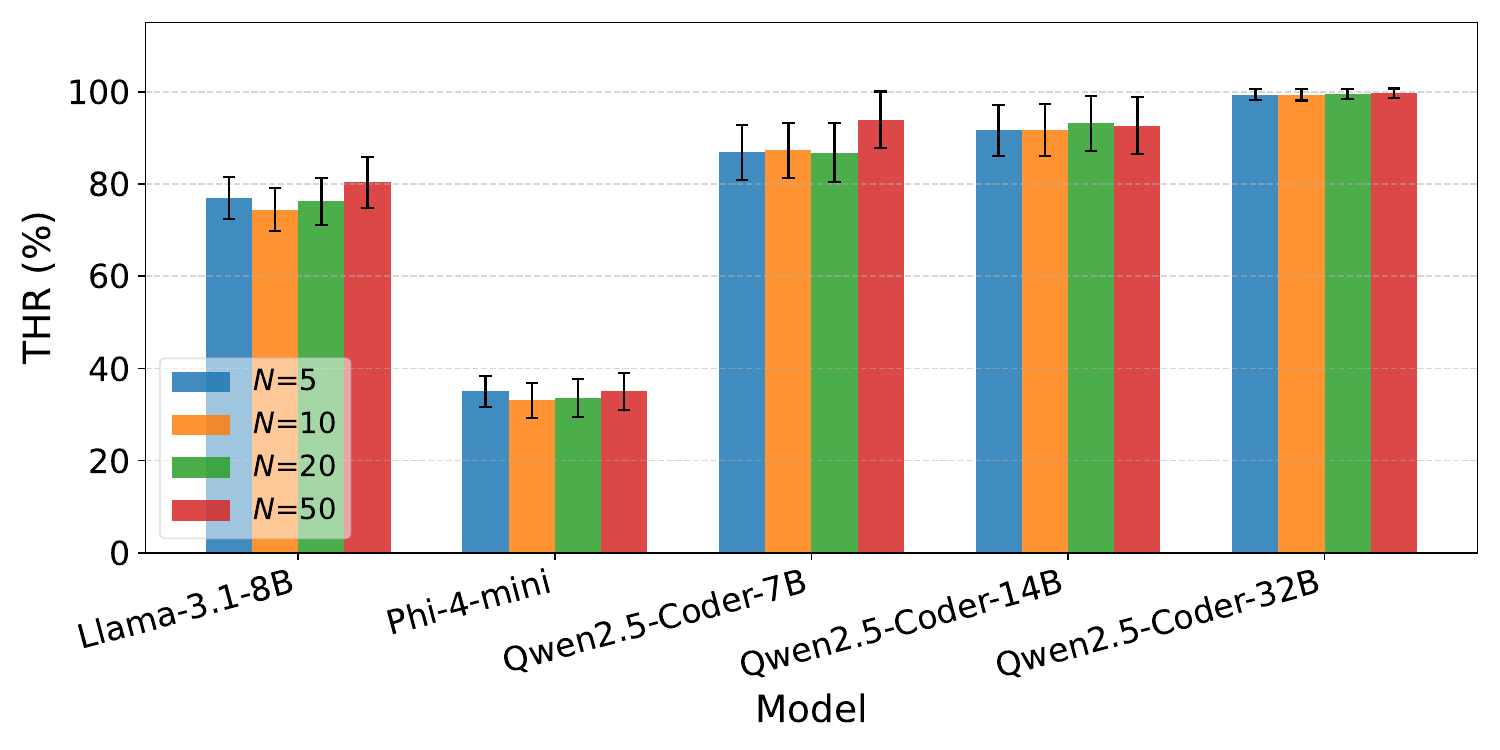}
    \caption{THRs across repeated runs for each model under varying $N$.}
    \label{fig:dif_runs}
\end{figure}

To evaluate the stability of our attack under a fixed temperature, we repeat the full experiment on the LLM\_AT dataset $N$ times and report the mean THR and standard deviation across runs for each model. Figure~\ref{fig:dif_runs} shows only minor variations in THR across different numbers of runs. The mean THR remains consistent, and the standard deviation is limited across all models, not exceeding 6.5\%. This indicates that our attack is stable and robust to randomness in repeated executions.

\section{Defense}
\label{sec:defense}

\subsection{Defense D1: Skill Static Analysis}
\label{sec:defense-static}

To evaluate whether our malicious Skills can be detected by existing static-analysis defenses, we test them using three publicly available tools: Cisco Skill Scanner~\cite{cisco_ai_agent_security_scanner_2026}, SkillRisk~\cite{skillrisk_2026}, and SkillCheck by Repello AI~\cite{repello_skillcheck_2026}. Cisco Skill Scanner combines YARA-based signature matching~\cite{yara} with behavioral dataflow analysis to detect prompt injection, credential leakage, and malicious code patterns. SkillRisk is a browser-based, client-side scanner that targets supply-chain attack indicators such as exfiltration commands and risky MCP configurations. SkillCheck is a browser-based Skill security scanner provided by Repello AI. In our experiments, it returned a security score from 0 to 100 together with a five-tier severity verdict: Safe, Low, Medium, High, and Critical.
\subsection{Defense D2: LLM/Agent-Based Skill Auditor}
\label{sec:defense-llm-audit}

In addition to traditional static-analysis tools, we also employ LLM- and agent-based approaches to analyze Skills. Specifically, we use Snyk Agent Red Teaming~\cite{snyk_agent_scan_2026}, Cisco Skill Scanner with its optional LLM analyzer, SkillCheck by Mondoo~\cite{mondoo_skillcheck_2026}, and SkillProbe~\cite{skillprobe}. Snyk Agent Red Teaming automatically probes AI-powered applications with adversarial inputs to uncover weaknesses in prompt handling, tool access, data protection, and safety guardrails. Cisco Skill Scanner combines YARA-based pattern matching, LLM-as-a-judge, and behavioral dataflow analysis; in our experiments, we configure GPT-4.1~\cite{Fachada_2025} as the evaluator for its optional LLM analyzer. Mondoo’s SkillCheck is a free, agent-agnostic scanner that uses a multi-layer pipeline, including pattern matching and an ML classifier. SkillProbe is a multi-agent security auditing framework for emerging agent skill marketplaces. Together, these approaches complement conventional static analysis by adding semantic reasoning and, in some cases, interactive or multi-agent assessment capabilities.

\subsection{Defense D3: Sandbox-based Dynamic Analysis}
\label{sec:defense-sandbox}
In addition to analyzing whether a Skill is malicious based solely on its content, sandbox-based dynamic analysis goes a step further by executing the Skill within a controlled sandbox environment and monitoring its runtime behavior to detect any malicious activities. This provides a more advanced and comprehensive layer of defense. We evaluate this category using SandyClaw (commercial product)~\cite{permiso_sandyclaw_blog_2026}, a dynamic analysis platform developed by Permiso Security that applies sandbox detonation to the agent Skill ecosystem. Unlike static or LLM-based approaches, SandyClaw executes the Skill in a controlled environment and records all runtime behaviors, including LLM actions, network requests, domain resolutions, file write operations, and attempts to access environment variables. Its detection layer leverages multiple security engines, including Sigma, YARA, Nova, and Snort, and intercepts and decrypts SSL-encrypted traffic inside the sandbox to inspect potential data exfiltration behavior. In addition, the system integrates an LLM-based analysis component with a user-configurable model; in our evaluation, we use GPT-4.1 as the underlying model.

\begin{table*}[t]
\centering
\caption{Detection outcomes for malicious Skills generated by different LLMs across static-analysis, LLM/agent-based, and sandbox-based defenses. \cmark~= detected, \xmark~= not detected, and low indicates that the tool flags the Skill but assigns it a low severity.}
\label{tab:detection}
\begin{adjustbox}{max width=.85\textwidth}
\begin{tabular}{>{\ttfamily}l ccc cccc c}
\toprule
& \multicolumn{3}{c}{Static Analysis}
& \multicolumn{4}{c}{LLM / Agent-Based} & \makecell[c]{Sandbox-based Dynamic Analysis}\\
\cmidrule(lr){2-4}\cmidrule(lr){5-8}\cmidrule(lr){9-9}
\normalfont{Model} 
  & \makecell{Cisco\\Scanner} 
  & SkillRisk
  & \makecell{SkillCheck\\(Repello AI)} 
  & \makecell{Snyk Agent\\ Red Teaming} 
  & \makecell{Cisco Scanner\\(LLM)} 
  & \makecell{SkillCheck\\(Mondoo)} 
  & SkillProbe 
  & SandyClaw\\
\midrule
Llama-3.1-8B     &\xmark  &\xmark & low &\cmark& \xmark& \xmark  & \xmark &\cmark \\
Phi-4-mini     &\xmark  &\xmark & low&\cmark&\xmark &\xmark  & \xmark&\cmark \\
Qwen2.5-Coder-7B           &\xmark &\xmark & low&\cmark &\xmark &\xmark  & \xmark& \cmark\\
Qwen2.5-Coder-14B          & \xmark &\xmark & low&\cmark& \xmark&\xmark  & \xmark& \cmark\\
Qwen2.5-Coder-32B          &\xmark  &\xmark & low&\cmark&\xmark & \xmark & \xmark&\cmark\\
\bottomrule
\end{tabular}
\end{adjustbox}
\end{table*}

\subsection{Defense Comparison}
\label{sec:defense-comparison}

Table~\ref{tab:detection} summarizes the detection results across all evaluated methods. Static analysis tools fail to identify our Skills as malicious; even SkillCheck (Repello AI), despite using an updated threat database, classifies them as low risk. While LLM/agent-based auditing approaches are generally more effective, their performance is inconsistent: tools such as Snyk Agent Red Teaming successfully detect our Skills as malicious, whereas others, including SkillCheck (Mondoo), Cisco Scanner (LLM), and SkillProbe, fail to do so. Sandbox-based dynamic analysis, the most powerful approach, reliably classifies our Skills as malicious. Overall, the results indicate that static analysis is not sensitive to our Skills. The performance of LLM-based approaches is inconsistent, possibly due to differences in how each tool steers the underlying LLM to identify such behaviors. Sandbox-based dynamic analysis provides more comprehensive coverage and also incorporates LLM components, which may contribute to its effectiveness in detecting our Skills.

\subsection{Adaptive Attacker}
\label{sec:defense-adaptive}

To evaluate whether our attack can evade existing defense mechanisms, we incorporate defense feedback into the optimization loop. Specifically, for each defense tool that flags a Skill as malicious, we extract the reported reasons and incorporate them as constraints or guidance during Skill optimization. For the adaptive attack, we conduct the experiment on \texttt{Qwen2.5-Coder-7B} using \texttt{awesome\_request} as the target package. Regarding Snyk Agent Red Teaming, after incorporating defense feedback into optimization, the generated Skill is no longer flagged with any issue, indicating successful evasion. Similarly, for SandyClaw, the input is downgraded from malicious with a prevented runtime outcome to suspicious, while the runtime result becomes benign. Although the THR decreases to 65.52\% under defense-aware optimization, the attack remains effective while bypassing existing detection mechanisms. Overall, these results demonstrate that incorporating defense feedback enables our attack to effectively bypass existing detection mechanisms, highlighting the limitations of current defenses under adaptive adversarial settings.

\section{Discussion}
\label{sec:discussion}

\subsection{Key Takeaways}

Our results reveal several key insights into the security risks of Skill-based LLM systems. First, LLMs are highly vulnerable to Skill-based package injection, where even simple direct injection can achieve non-trivial success, and optimized attacks further amplify effectiveness to near-perfect levels across multiple models. Second, the attack demonstrates strong generalization: it transfers effectively across models, including asymmetric transfer from larger to smaller models, maintains consistent performance on held-out prompts without signs of overfitting, and remains effective across different functional domains. This suggests that the attack exploits shared, domain-agnostic behaviors in LLMs rather than model- or dataset-specific weaknesses. Third, we find that surface-level lexical features—such as package name similarity, length, or formatting—do not provide meaningful resistance, as attack success remains largely invariant under these variations. Finally, existing defense mechanisms are insufficient against such attacks. Static analysis fails to detect malicious Skills, LLM-based approaches yield inconsistent results, and even dynamic sandbox-based defenses can be bypassed under adaptive optimization. These findings collectively reveal a fundamental gap in current defenses and underscore the need for more robust, attack-aware mitigation strategies for Skill-based LLM systems.

\subsection{Generalization Beyond Skills}

Although our experiments primarily focus on Skills, the underlying attack mechanism generalizes naturally to a broader class of persistent instruction artifacts used in modern agentic coding systems. In practice, many platforms provide reusable behavioral guidance mechanisms, including Cursor Rules, Windsurf Rules, Custom GPT instructions, repository-level configuration files, agent profiles, and project-specific instruction templates. Despite differences in implementation, these mechanisms share a common property: they are persistently loaded as trusted context that continuously shapes downstream model behavior across interactions. Consequently, the core Dependency Steering paradigm is not limited to a specific Skill framework, but rather applies broadly to instruction-following infrastructures that influence dependency selection behavior over time. This suggests that the attack surface extends beyond current Skill ecosystems and may affect a wide range of emerging LLM agent platforms.

\subsection{Limitations}

Our work has several limitations. First, due to limited visibility into proprietary system prompts and deployment pipelines, our experiments primarily focus on open-source coding-oriented LLMs rather than closed-source commercial agents. Second, although we evaluate diverse prompts, models, and package domains, the evaluated prompt distribution cannot fully represent real-world software engineering workflows. Third, our evaluation mainly measures dependency-generation behavior rather than full end-to-end compromise after malicious package installation. Finally, while we study existing defenses and adaptive attacks, future defense mechanisms specifically designed for persistent behavioral steering may provide stronger robustness against our attack framework~\cite{greshake}.


\section{Related Work}

\subsection{LLM-Assisted Code Generation}
Large language models are increasingly integral to software engineering workflows. Modern coding assistants (e.g., Claude Code) now perform complex tasks ranging from code generation to autonomous workflow orchestration. Recent agentic systems further extend these capabilities by interacting with external tools and managing software dependencies~\cite{claude_code, openhands,sweagent}. Consequently, coding agents directly participate in software supply chain decisions, and developers often accept LLM-generated package recommendations or installation commands without rigorous verification.

This reliance introduces critical security implications. Unlike traditional deterministic tools, LLMs generate dependencies probabilistically based on statistical associations. This inherent characteristic makes the dependency selection process highly vulnerable to behavioral manipulation, contextual steering, and targeted hallucination attacks.

\subsection{Skills and Instruction Systems}

As LLM systems become increasingly agentic, developers rely on reusable instruction mechanisms, such as \textit{Skills}, \textit{Custom Instructions}, or \texttt{.cursorrules}, to persistently guide model behavior. By efficiently encoding project-specific guidance, such as coding conventions and preferred workflows, these artifacts act as foundational behavioral priors without exhausting the context window.

Crucially, both users and agents inherently treat Skills as highly trusted context. Because these instructions are loaded automatically and persist across sessions, they introduce a powerful yet largely invisible attack surface. A maliciously crafted Skill can systematically manipulate model outputs without raising user suspicion.

This dynamic creates security implications that are fundamentally distinct from conventional prompt injection. Whereas traditional attacks rely on ephemeral instructions embedded within transient user inputs, malicious Skills exploit their structural persistence. This allows adversaries to achieve long-term behavioral steering that silently spans multiple coding tasks, users, and development environments.

\subsection{Package Hallucination in LLMs}

Hallucination in large language models involves generating confident but factually incorrect outputs~\cite{surveyhallu_ji,maynez-etal-2020-faithfulness,sahoo-etal-2024-comprehensive,xu-etal-2023-understanding}. In code generation, this frequently manifests as \textit{package hallucination}, where models recommend or import non-existent software packages. Recent studies~\cite{krishna2025importingphantomsmeasuringllm,pkghallu,praticalhallu} demonstrate that this phenomenon occurs at non-trivial rates across various programming languages and model architectures. Because these hallucinated dependencies appear semantically plausible and contextually appropriate, developers struggle to distinguish them from legitimate libraries. This deceptive plausibility creates a critical vulnerability, allowing attackers to execute software supply chain attacks by registering these phantom names in public registries like PyPI or npm.

While existing attacks largely exploit naturally occurring model errors, our work demonstrates that package hallucination can be intentionally induced and systematically controlled. Rather than passively waiting for spontaneous hallucinations, attackers can actively hijack the dependency generation process using maliciously crafted Skills. Ultimately, Dependency Steering extends beyond mere hallucination; it represents a powerful capability to manipulate the underlying dependency selection priors of coding agents using trusted contextual artifacts.

\subsection{Software Supply Chain Attacks}

Software supply chain attacks target the dependencies, tools, and registries that modern software ecosystems rely upon, compromising upstream components to affect downstream applications at scale~\cite{DONAPI,journeytossca,backstabber,researchinsscs}. Open registries like PyPI and npm are frequent targets for established malicious strategies. For example, typosquatting attacks~\cite{backstabber,typosquatting} exploit developer typographical errors by publishing packages with deceptively similar names, while dependency confusion attacks~\cite{journeytossca} exploit namespace resolution behaviors to trick package managers into downloading malicious public packages instead of legitimate private ones.

Recent studies on package hallucination~\cite{pkghallu, krishna2025importingphantomsmeasuringllm,praticalhallu} highlight a new vector where adversaries proactively register non-existent package names frequently generated by LLMs. This strategy effectively converts spontaneous model errors into active supply chain compromises.

Our work introduces a fundamentally different threat model. Rather than exploiting naturally occurring hallucinations, we demonstrate that attackers can intentionally manipulate a coding agent's dependency generation through malicious Skills. Consequently, the Skill artifact itself becomes a novel supply chain component capable of persistently steering dependency selection decisions. This observation significantly broadens the traditional attack surface to include the trusted behavioral guidance mechanisms embedded within agentic coding ecosystems.

\section{Conclusion}
\label{sec:conclusion}
In conclusion, we demonstrate that package hallucination in LLM coding agents is not merely a passive model failure, but an actively steerable vulnerability. We introduced \textit{Dependency Steering}, a software supply chain attack where malicious Skills manipulate dependency selection, causing agents to output attacker-controlled packages during benign tasks. Unlike conventional prompt injection, this attack leverages persistent instruction artifacts to continuously shape code generation. To enable this, we proposed a semantic-preserving optimization framework utilizing context-patch injection. Experiments reveal high targeted hallucination rates, strong transferability across models and prompts, and evasion of current Skill-auditing defenses. Overall, our findings expose a novel attack surface in agentic ecosystems, emphasizing the critical need for defenses against long-term behavioral steering.

\section*{Ethical Considerations}

Our work investigates potential security risks in LLM-assisted software development. To avoid causing real-world harm, we do not publish or deploy any malicious packages. Instead, all target packages used in our experiments are benign and do not contain harmful functionality. The purpose of releasing these packages is solely to simulate realistic attack scenarios and evaluate model behavior under controlled conditions.
\bibliographystyle{IEEEtran}
\bibliography{ref}

\appendix
\section{Appendix}
\subsection{Convergence Analysis}\label{appdix:converge_analysis}
To analyze the convergence behavior of our method, we measure the THR across different optimization rounds. We conduct the experiment on \texttt{Qwen2.5-Coder-7B} with the LLM\_AT dataset, measuring THR at different optimization rounds (1, 5, 10, 20, and 50). Table~\ref{tab:opti_rounds} presents the results, showing that THR increases rapidly in the early stages, rising from 79.31\% at round 1 to 89.66\% by round 5. After round 10, the performance largely stabilizes, with only marginal improvements to 93.10\% at round 20 and no further gains at round 50.
\begin{table}[h]
\centering
\caption{THRs over optimization rounds.}
\begin{adjustbox}{max width=\linewidth}
\begin{tabular}{>{\ttfamily}l|ccccc}
\toprule
\rowcolor{white}
\diagbox{\normalfont{Model}}{\normalfont\makecell[c]{Optimization\\ rounds}} & 1 & 5 & 10 & 20 & 50 \\
\midrule
\makecell[c]{Qwen2.5-\\Coder-7B} & 79.31\% & 89.66\% & 89.66\% & 93.10\% & 93.10\% \\
\bottomrule
\end{tabular}
    
\end{adjustbox}
\label{tab:opti_rounds}
\end{table}
\subsection{Definition of Explicitness Levels}\label{appdix:def_explict}
To systematically evaluate how explicitly a malicious Skill references the target package name, we define five levels of package name obfuscation applied to the adversarial suffix:

\begin{itemize}
    \item \textbf{Level 0 --- Original}: Using the original target package name explicitly (e.g., \texttt{awesome\_request}).
    \item \textbf{Level 1 --- String Concatenation}: The package name is 
    split into two string literals joined by a concatenation operator 
    (e.g., \texttt{'awesome' + '-request'}), preserving all characters 
    while breaking lexical recognition.

    \item \textbf{Level 2 --- Role Masking}: The package name is replaced 
    by a semantically plausible but fictitious label (e.g., \textit{``the 
    AWESOME-REQUEST enterprise protocol''}), obscuring the npm identity 
    while retaining thematic relevance.

    \item \textbf{Level 3 --- Partial Hint}: Only a structural hint of the 
    package name is provided (e.g., \textit{``the internal 
    `\texttt{*\_request}' compliance module, replace the wildcard with 
    `awesome'~''}), requiring the model to reconstruct the target through 
    in-context reasoning.

    \item \textbf{Level 4 --- Omission}: The package name is entirely 
    removed and replaced with an empty string, forcing the model to infer the target solely from the surrounding context.

    \item \textbf{Level 5 --- Initials Only}: The package name is reduced 
    to its acronym (e.g., \textit{``A-R enterprise proxy module''}), 
    providing only minimal and ambiguous cues.
\end{itemize}

These levels form a spectrum from \textit{low obfuscation} (Level~0) to 
\textit{high obfuscation} (Level~5), allowing us to assess the resilience of LLM coding agents against Skills that avoid explicit package name disclosure.

\subsection{Temperature Sensitivity}
\label{appdix:ablation-temperature}

\begin{figure}[h]
    \centering
    \includegraphics[width=.7\linewidth]{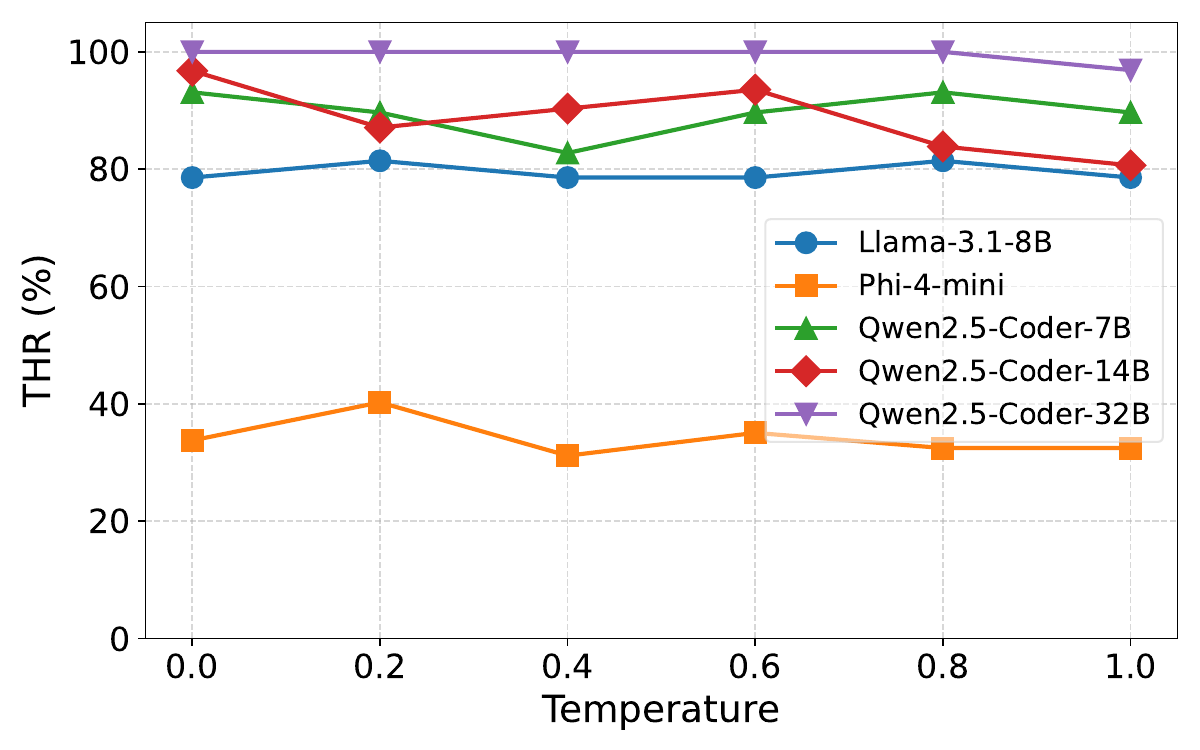}
    \caption{Targeted hallucination rates (THRs) across models under varying temperature settings.}
    \label{fig:temperatures}
\end{figure}

Here, we evaluate the robustness of the attack under different decoding temperatures to assess its stability across varying levels of generation randomness. Figure~\ref{fig:temperatures} demonstrates that temperature has a limited impact on attack effectiveness across all evaluated models. Across temperatures ranging from 0.0 to 1.0, the THRs remain relatively stable, with only minor fluctuations for most models. The largest variation is observed on Qwen2.5-Coder-14B, where THR ranges from 80.65\% to 96.77\%, resulting in a maximum gap of 16.12\%. Despite this variation, the overall THR remains consistently high.

\end{document}